# Picturing the Gap Between the Performance and US-DOE's Hydrogen Storage Target: A Data-Driven Model for MgH$_2$ Dehydrogenation

Chaoqun Li[a], Weijie Yang[a*], Hao Liu[a], Xinyuan Liu[a], Xiujing Xing[b], Zhengyang Gao[a], Shuai Dong[a], and Hao Li[c*]

[a] School of Energy and Power Engineering, North China Electric Power University, Baoding, 071003, Hebei, China

[b] Chemistry Department, University of California, Davis 95616, United States

[c] Advanced Institute for Materials Research (WPI-AIMR), Tohoku University, Sendai 980-8577, Japan

\* Corresponding authors:

Prof. Weijie Yang (yangwj@ncepu.edu.cn);

Prof. Hao Li (li.hao.b8@tohoku.ac.jp).

# Abstract

Developing solid-state hydrogen storage materials is as pressing as ever, which requires a comprehensive understanding of the dehydrogenation chemistry of a solid-state hydride. Transition state search and kinetics calculations are essential to understanding and designing high-performance solid-state hydrogen storage materials by filling in the knowledge gap that current experimental techniques cannot measure. However, the *ab initio* analysis of these processes is computationally expensive and time-consuming. Searching for descriptors to accurately predict the energy barrier is urgently needed, to accelerate the prediction of hydrogen storage material properties and identify the opportunities and challenges in this field. Herein, we develop a data-driven model to describe and predict the dehydrogenation barriers of a typical solid-state hydrogen storage material, magnesium hydride ($MgH_2$), based on the combination of the crystal Hamilton population orbital of Mg-H bond and the distance between atomic hydrogen. By deriving the distance energy ratio, this model elucidates the key chemistry of the reaction kinetics. All the parameters in this model can be directly calculated with significantly less computational cost than conventional transition state search, so that the dehydrogenation performance of hydrogen storage materials can be predicted efficiently. Finally, we found that this model leads to excellent agreement with typical experimental measurements reported to date and provides clear design guidelines on how to propel the performance of $MgH_2$ closer to the target set by the United States Department of Energy (US-DOE).

***Keywords***: Solid-state hydrogen storage, $MgH_2$, dehydrogenation kinetics, crystal orbital Hamilton population (COHP), performance descriptor.

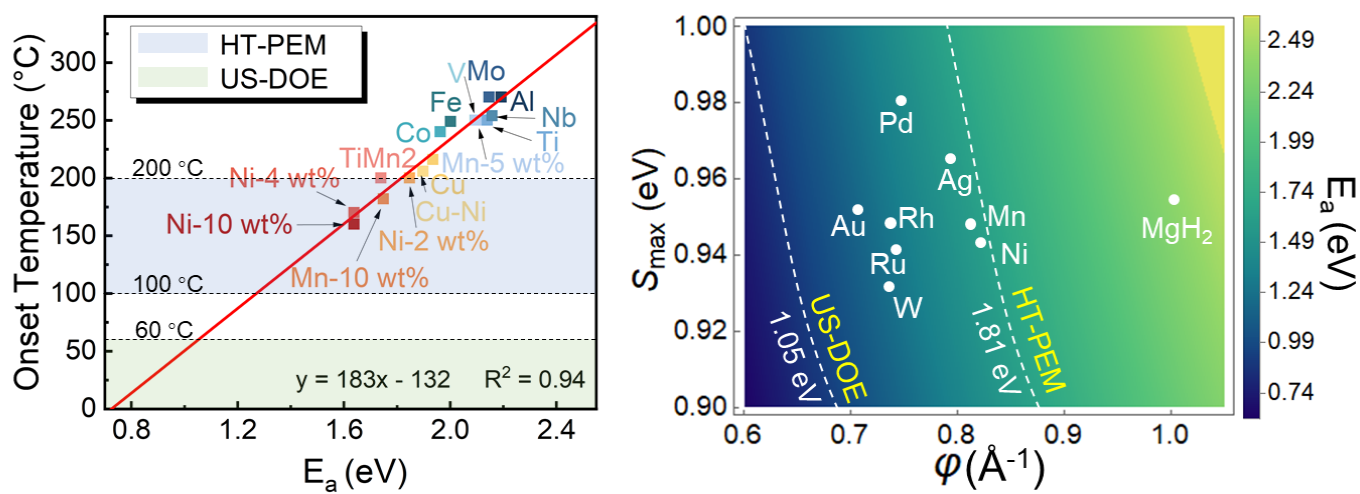

**Graphical Abstract**

# 1. Introduction

Utilizing hydrogen energy is the key to the sustainable future.[1-3] In recent years, solid-state hydrogen storage has emerged as a prominent research focus driven by its emphasis on safety and efficiency, as hydrogen storage research continues to advance.[4-6] Magnesium hydride ($MgH_2$) has become a key research object in the hydrogen storage industry due to its excellent hydrogen storage capacity and reversibility.[7, 8] $MgH_2$ exhibits a high theoretical hydrogen storage capacity of 7.66 wt%, and the abundance of Mg on the Earth makes $MgH_2$ an optimal choice for sustainable hydrogen storage. Moreover, Mg is an inexpensive element, enhancing the cost-effectiveness of $MgH_2$ and positioning it as an excellent candidate for commercial applications. Due to these unique features, this material is regarded as one of the most promising candidates for solid-state hydrogen fuel cells.[9]

However, significant improvements are still needed for Mg-based solid hydrogen storage materials before they can be broadly used in industry, with their sluggish dehydrogenation kinetics and thermodynamics as the key bottlenecks. At present, scientists have developed a variety of methods such as nanofabrication,[10] alloying,[11] and catalyst doping[12] to facilitate the dehydrogenation process of $MgH_2$. With the advancement of relevant science and technology, *ab initio* calculations (primarily based on density functional theory, DFT) have been employed to calculate the dehydrogenation barriers and reaction enthalpies of $MgH_2$, to accelerate the design and understanding of Mg-based solid hydrogen storage materials.

The hydrogen storage and release properties of hydrides will be dominated by the dehydrogenation kinetics when the crystalline phase structure change of the material itself is not considered. According to the Arrhenius formula,[13] the dehydrogenation barrier of a solid-state hydride is an important indicator to evaluate the difficulty in the dehydrogenation process of a hydrogen storage material. The breaking and formation of chemical bonds are ubiquitous in this process, and the energy barrier is crucial to determine the underlying mechanism. Therefore, finding the physical and chemical determinants of dehydrogenation barriers has always been a primary objective. Recently, combined with DFT calculations, *ab initio* molecular dynamics (AIMD), and electronic structure analyses, an interesting "burst effect" was identified during dehydrogenation on a most common $MgH_2$(110) surface, showing that the surface-layer dehydrogenation of $MgH_2$ is the key bottleneck due to the most sluggish dehydrogenation kinetics over a $MgH_2$ structure.[14-17] Similarly, it was concluded by experiments that surface modification of $Mg/MgH_2$ is crucial for accelerating the recombination of hydrogen.[18, 19] Promoting surface dehydrogenation by engineering strategies can lead to a significant improvement in the overall hydrogen release performance of $MgH_2$.[20-23] Therefore, understanding the dehydrogenation behavior in the surface and near-surface layers is essential in designing a more efficient

$MgH_2$ for hydrogen storage, while the dehydrogenation and hydrogen diffusion in the inner layers are more facile.

Although the accuracy and efficiency of transition state search methods from nudged elastic band (NEB)[24] to climbing-image NEB (CI-NEB)[25] and their derivatives[26] have been improved, they are relatively time-consuming for locating an energy barrier. Besides, NEBs can only capture one transition state for each calculation. Though other transition state search methods (*e.g.*, the Dimer method[27]) and AIMD simulations (*e.g.*, metadynamics[28] and high-temperature MD simulations) can sample the potential energy surface in a wider range, their ultra-high computational cost is the critical bottleneck for solid-state hydrogen storage materials. Therefore, it is particularly important to develop a strategy to directly predict the dehydrogenation barrier, *e.g.*, by correlating much less computationally expensive properties with the energy barrier. This will not only help identify the basic determinants of the dehydrogenation performance of a solid-state hydride, but also pave an avenue to effectively design high-performance materials. Intuitively speaking, the energy barrier should be a function of the geometric and electronic characteristics of the reaction domains, where the direct cause of energy barrier is the energy change during the processes of breaking and forming chemical bonds.

Herein, we studied the chemical, structural, and electronic properties of the dehydrogenation processes of $MgH_2$, with the goal of unraveling the determinants of the dehydrogenation barrier change of different surface structures. Due to the "burst effect" of $MgH_2$ dehydrogenation,[14-17] we aim to build a model to precisely describe $MgH_2$ dehydrogenation for the surface and near-surface domains. Through a data-driven strategy, the correlation between dehydrogenation energy barrier and dehydrogenation descriptors has been progressively improved through the gradual incorporation of the crystal orbital Hamiltonian population (COHP),[29, 30] distribution characteristics of H vacancies, atomic H displacement, and metal doping. Among them, COHP is a "bond-weighted" density of states between a pair of adjacent atoms. A COHP diagram indicates the bonding and antibonding contributions to the band-structure energy. The ICOHP is the integrated value of COHP below the Fermi level, which can be understood as the energy contribution of bonding electrons shared between two atoms, and can accurately reflect the magnitude of bond strength.[31, 32] In this study, we report the first direct model that unravels the essential insights into the properties of $MgH_2$ dehydrogenation. Moreover, descriptors can help screen potential materials, requiring only experimental verification of candidate structures to select novel doped materials. This approach significantly reduces the material design cycle and research costs. Finally, the computed values from this model agree with the temperature variation observed in many typical experimental measurements reported to date, and provide new design guidelines on how to propel the performance of $MgH_2$ closer to the target set by the United States Department of Energy (US-DOE).

## 2. Results and Discussion

### 2.1. Foundation of the Model

Identifying the key descriptors requires prior data of MgH$_2$ dehydrogenation, including those from multiple dehydrogenation paths and structures. These data have been obtained from the calculations at the same level of accuracy in our current and previous studies,[14] including the energy barrier changes for the sequential dehydrogenation of MgH$_2$(110). The important factors influencing the dehydrogenation of MgH$_2$ include structural changes and electronic distribution, as shown in **Figure 1a**.

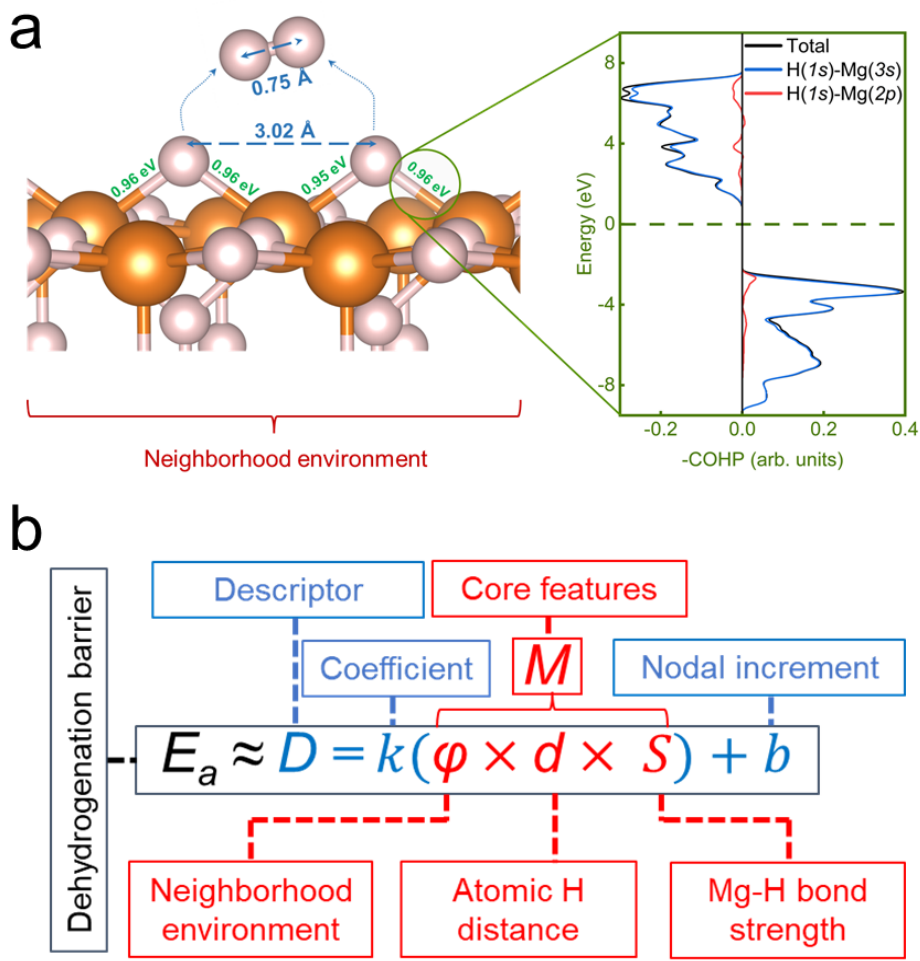

**Figure 1** The composition of descriptors for the kinetics of MgH$_2$ dehydrogenation. (a) A typical example of MgH$_2$ dehydrogenation and its bonding state distribution. White and orange spheres represent H and Mg, respectively. (b) Descriptors that may affect the MgH$_2$ dehydrogenation kinetics, where $E_a$ represents the dehydrogenation energy barrier of MgH$_2$, and $D$ represents the descriptor of energy barrier. $\varphi$, $d$, and $S$ represent the parameters composed of the neighboring environment, H-H distance, and Mg-H bonding strength, respectively. $M$ represents the part of the model that contains basic parameters (*i.e.*, $\varphi$, $d$, and $S$). $k$ and $b$ are the coefficients and intercepts that linearly connect with $M$ and $D$, respectively.

In addition, a positive correlation is found between the Mg-H bonding strength and the dehydrogenation barrier.[14] Therefore, the integral value of the crystal orbital Hamilton population (ICOHP) is recognized as one of the main considerations in model construction. During $H_2$ formation, the cleavage of Mg-H bonds and H-H bond formation are the main reasons for forming dehydrogenation barriers. Therefore, in addition to the Mg-H bonding strengths, the timing of the H-H bond formation should also be considered. The formation of H-H bonds can prevent the increase in the energy generated by bond-breaking, *i.e.*, the increase in energy barriers. Moreover, the farther away the two atomic H are, the lower the collision probability would be, and therefore the required activation energy for the reaction will increase. Because of this, there is also a positive correlation between the distance of the two atomic H and the dehydrogenation barrier. Changes in H-H distance and bond strengths during the ten previously analyzed dehydrogenation reactions[14] exhibit a consistent trend with the variation in the dehydrogenation barrier, and both factors are easily obtainable. In addition to these two factors, differences in the numbers and types of atoms in the neighboring environment can also affect the reaction energy barrier. Therefore, through the consideration of these three factors, a foundation of the model can be established, as illustrated in **Figure 1b**.

Next, we build the formula of $M$ as the function of the descriptors with parameter refinements. The initial expression of $M$ is shown in **Eq. 1**:

$$M = \varphi \times d_{H-H} \times S \tag{1}$$

where $\varphi$ is the pre-factor generated due to different neighboring environments, $d_{H\_H}$ is the distance between the atomic H participating in $H_2$ formation, and $S$ is the Mg-H bonding strength variable composed from -ICOHP. As bond breaking is an endothermic process, the timing of bond cleavage will affect the energy barrier. Due to the proximity of two hydrogen atoms to each other during the dehydrogenation process of $MgH_2$, the Mg-H bonds on the two sides will break first. The Mg-H bonds on the two sides are defined as the "outside" of the Mg-H bonds, while the Mg-H bonds between the two atomic H are referred to be the "inside" of Mg-H bonds (note: definitions of "inside" and "outside" of a reaction are illustrated in **Figure S1**). Therefore, $S$ can be considered in the following four parts:

i) Due to the preferential breakage of the bonds outside the atomic motion direction of the reaction, the average value of -ICOHP for all Mg-H bonds from only outside the reaction direction is considered (**Eq. 2**):

$$S_{out} = \frac{\sum S_{H1\_out} + \sum S_{H2\_out}}{n} \tag{2}$$

ii) Consider only the average -ICOHP of the Mg-H bonds inside the reaction direction (**Eq. 3**):

$$S_{in} = \frac{\sum S_{H1\_in} + \sum S_{H2\_in}}{n} \quad (3)$$

iii) Consider the average -ICOHP of all Mg-H bonding involved in the reaction (**Eq. 4**):

$$S_{ave} = \frac{\sum S_{H1\_in} + \sum S_{H1\_out} + \sum S_{H2\_in} + \sum S_{H2\_out}}{n} \quad (4)$$

iv) For each atomic H, only one Mg-H bond with the strongest strength is considered, and the average value is taken in **Eq. 5**:

$$S_{max} = \frac{max(S_{H1\_out}, S_{H1\_in}) + max(S_{H2\_out}, S_{H2\_in})}{2} \quad (5)$$

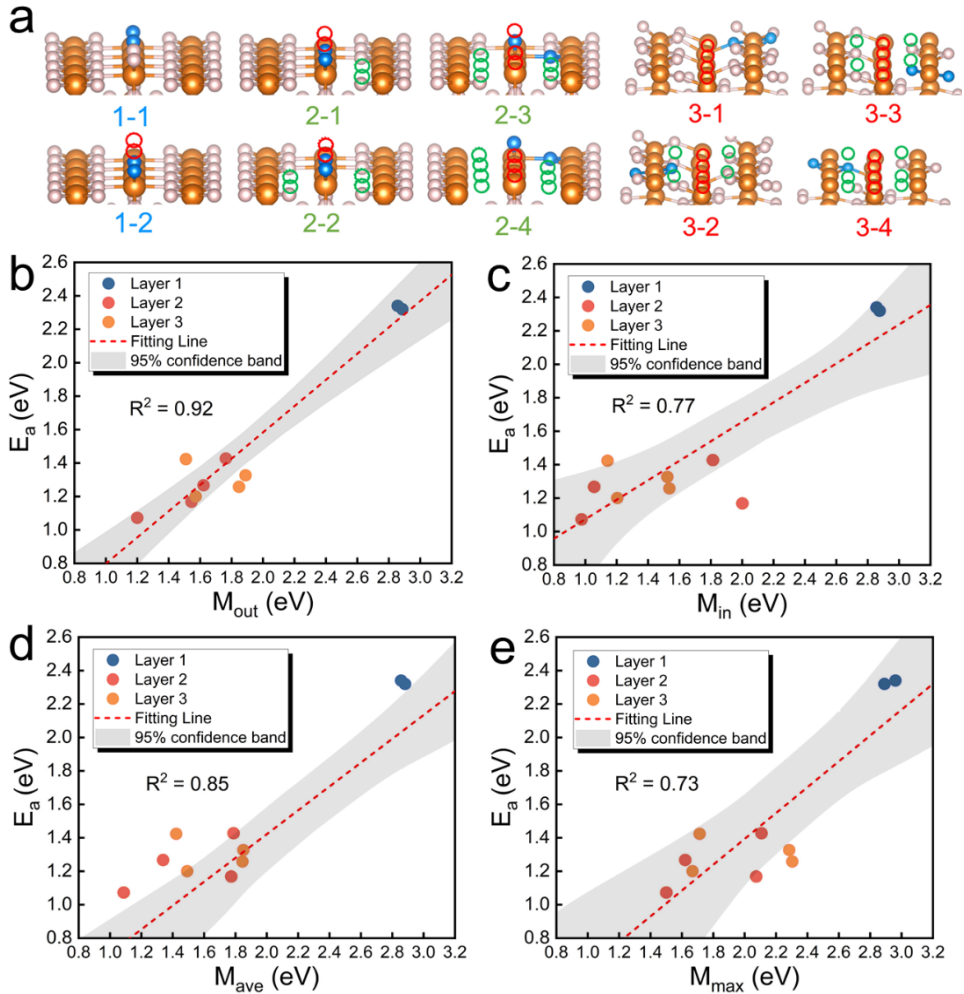

**Figure 2** The initial exploration of the descriptors for MgH$_2$ dehydrogenation barrier. (a) MgH$_2$ dehydrogenation structures for constructing the initial model. White and orange spheres represent H and Mg, respectively. Blue spheres and dashed circles represent the atomic H involved in the reaction and H vacancy, respectively. (b-e) Correlation analysis of the initial model considering neighboring environments. Correlations between the energy barriers and model values using (b) $S_{out}$, (c) $S_{in}$, (d) $S_{ave}$, and (e) $S_{max}$, respectively.

The starting data for model construction uses the DFT-calculated results of sequential dehydrogenation.[14] The energetics consider the correction of zero-point energies to ensure that the transition state energy is precisely mapped into the descriptor. Related structures of the reactions are shown in **Figure 2a**. Next, the initial parameters of the model and energy barrier are obtained by counting the distance of atomic H before the reaction and the corresponding -ICOHP in the reaction direction (**Table S1**). Note that neglecting the neighborhood coefficient would lead to a weak correlation between the model and energy barrier (**Figure S2** and **Table S2**). The -ICOHP values of the $MgH_2$ initial reaction structure are higher for the desorption of second-layer hydrogen compared to those of the third layer. When the descriptors are calculated without considering neighborhood coefficients, the energy barrier for the desorption of second-layer hydrogen is higher than that for the desorption of the third-layer hydrogen. However, DFT calculations reveal that energy barriers for the desorption of the second- and third-layer hydrogen should be similar. Therefore, to establish a linear correlation between the descriptor and dehydrogenation barrier, neighborhood coefficients need to be considered.

To explore potential correlations more effectively, the neighborhood coefficients of dehydrogenation in the second and third layers are adjusted (**Figure 2b-e** and **Tables S1, S3**). Interestingly, the dehydrogenation barrier has a significant linear relationship with *M* only when the average value of outer bond strength is used as *S*. Nevertheless, other representations of bond strength may also have the potential to accurately describe the energy barriers. The presence of H vacancies and the type of atoms are the important factors affecting the electronic distribution in $MgH_2$. Therefore, it is necessary to augment the descriptor's neighborhood coefficients by introducing more dehydrogenation data with varying H vacancy types and electronegativities.

**2.2. Effects of H Vacancy**

Next, five new sets of dehydrogenation data are added to further explore the relationship between H vacancies and dehydrogenation energy barriers (**Table S4** and **Figure S3**). The considered structures are shown in **Figure 3a**. In addition, the region of neighboring positions where the presence of a hydrogen vacancy can affect dehydrogenation is also defined, and the types of H vacancies are classified (**Figure 3b**). Based on the neighborhood coefficient of the first assignment (**Table S1**), the coefficient is adjusted by the change of H vacancies. To adjust properly, equal neighborhood coefficients should be given to both types of H vacancies when their numbers are equal. For a specific type of H vacancies, if its number increases monotonically, the neighborhood coefficient should change monotonically. The linear correlation between the model and the energy barrier after adjusting the neighborhood coefficient should be more significant. After the dehydrogenation paths are analyzed, the number of H vacancies for each path is counted. Subsequently, we summarize the neighborhood coefficient formula for H vacancies.

After modifying the neighborhood coefficients based on previous analysis (**Table S5**), a relationship between the dehydrogenation barrier and the model for modifying the neighborhood coefficients is established (**Figure 3c-f** and **Table S6**). Note that it is difficult to achieve a significant correlation with a high $R^2$ when using the average value of the outer bond strength as $S$, regardless of how the neighborhood coefficient is adjusted. Moreover, we observed that the model exhibits a significant correlation with the energy barrier when using the average value of all bonding strengths and the two strongest bonds on either side of the reaction direction as $S$, and the vacancy neighborhood coefficient satisfies **Eqs. 6-7**:

$$\varphi_H = 1 + 0.125k_1 - 0.125k_2, \quad k_1 < 2 \tag{6}$$

$$\varphi_H = 1 + 0.125k_1 + 0.125k_2, \quad k_1 \geq 2 \tag{7}$$

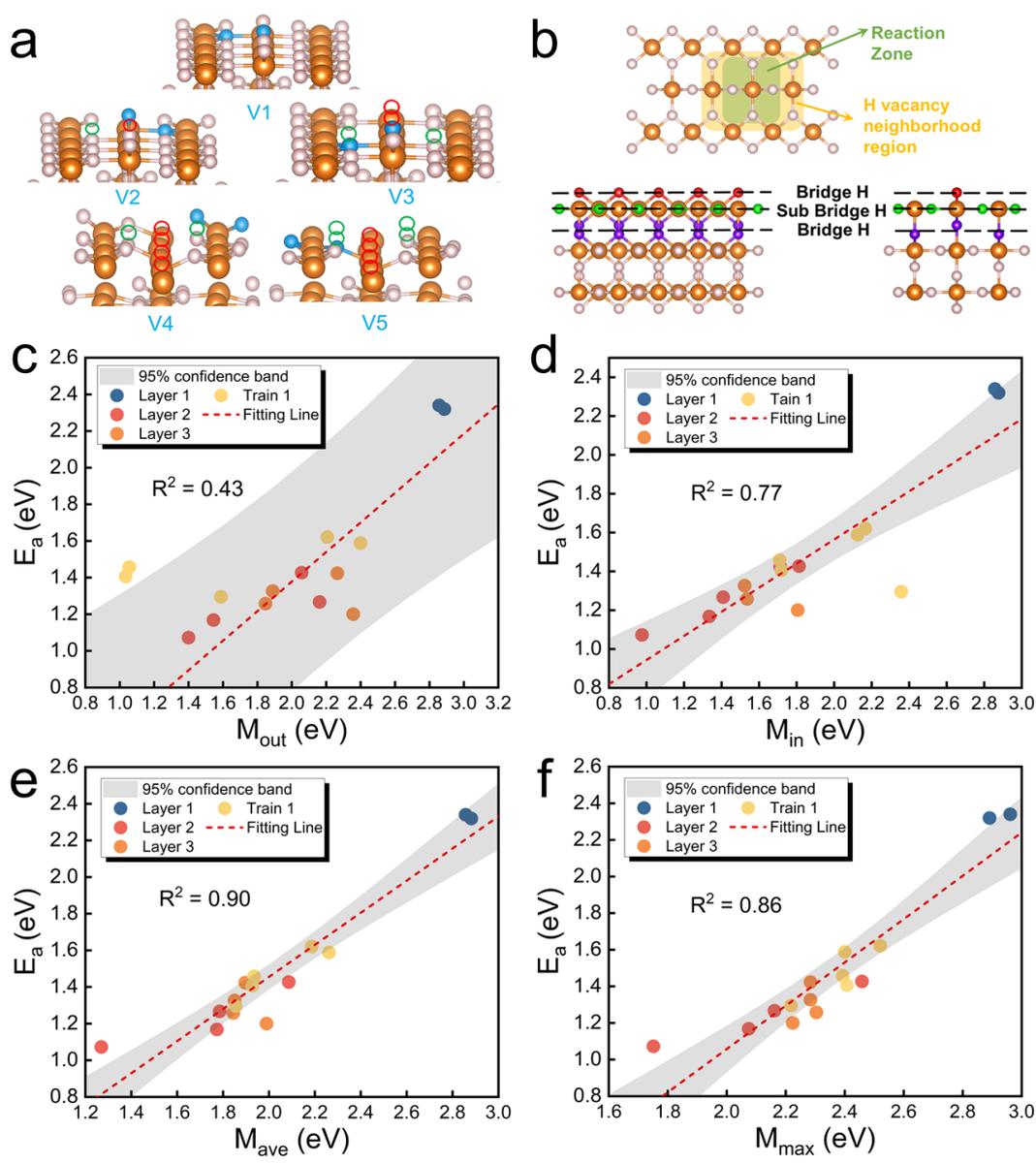

**Figure 3** Exploration of the region where H vacancies capable of influencing the $MgH_2$ dehydrogenation barrier are located and model correlation analysis to modify the neighborhood coefficients of H defect. (a)

Optimized structures of MgH$_2$ dehydrogenation to explore the effects of H defects. White and orange spheres represent H and Mg, respectively. Blue spheres represent the atomic H involved in the reaction. (b) Schematic illustration of the neighborhood range of atomic H occupation and H defect that affect the dehydrogenation. White and orange spheres represent H and Mg, respectively. Red, green, and purple spheres represent the first, second, and third layers of atomic H, respectively. (c-f) Correlations between energy barriers and model values using (c) $S_{out}$, (d) $S_{in}$, (e) $S_{ave}$, and (f) $S_{max}$, respectively.

Where $k_1$ and $k_2$ respectively represent the number of bridge H and sub-bridge H vacancies in the initial structure under different dehydrogenation paths. When there are two bridge H vacancies in the dehydrogenation reaction region, the presence of both bridge H vacancies and sub-bridge H vacancies will increase the coefficient for vacancy neighborhood. However, if there is a bridge H presenting in the reaction region, the coefficient for the vacancy neighborhood will diminish with the increase of sub-bridge H vacancies. This implies that the presence of bridging H will determine whether sub-bridge H vacancies would promote or hinder dehydrogenation.

## 2.3. Distance Between Two Atomic H

Our analysis reveals that the two H atoms can be viewed as "the endpoints of a spring", whereas the strength of the Mg-H bond represents the magnitude of the pulling force exerted on both ends (**Figure S4**). At the macro level, force and energy are mainly calculated around changes in the spring length as the Hooke's law shown in **Eq. 8**. Therefore, $d_{H-H}$ should be approximated by the distance difference, $\Delta x$. It is known that the distance between the two atoms in H$_2$ ($d_{H2}$) is 0.75 Å. Consequently, distance information ($d_{H-H}$) is transformed by taking the difference between the atom spacing of H in MgH$_2$ and that of H in H$_2$. This attempt yields remarkable findings, whereby the correlation analysis of various $S$ models demonstrates a significant improvement (**Figure 4a-d** and **Table S7**). As such, the atomic distance information is modified using **Eq. 9**:

$$F = k\Delta x \qquad (8)$$

$$d = d_{H-H} - d_{H_2} \qquad (9)$$

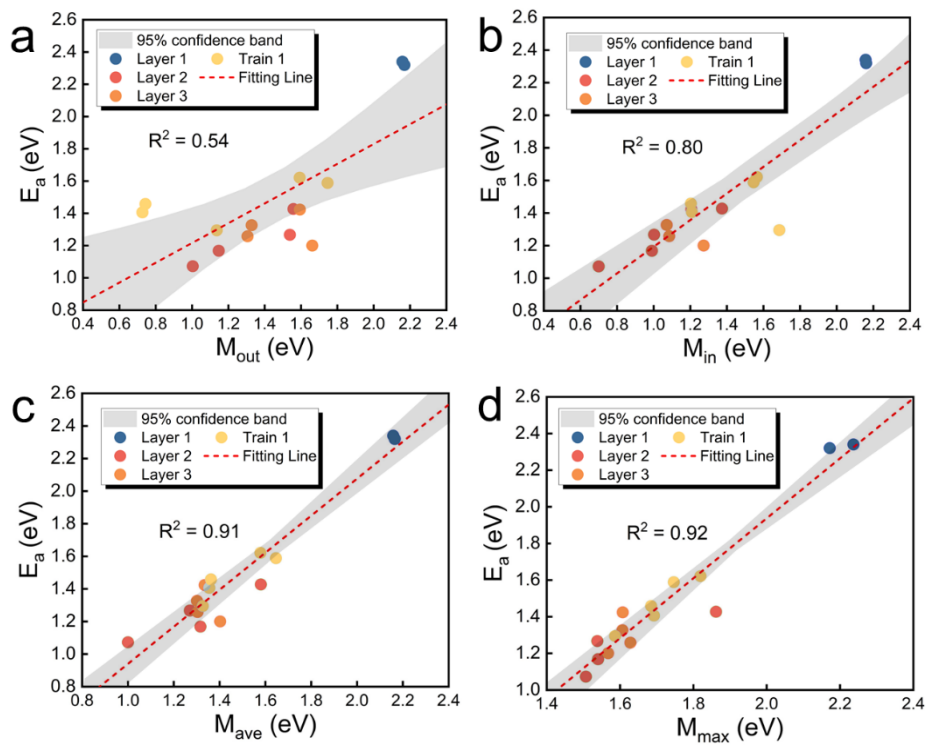

**Figure 4** Model correlation analysis to modify the distance information. (a-d) Correlation between energy barriers and model values using (a) $S_{out}$, (b) $S_{in}$, (c) $S_{ave}$, and (d) $S_{max}$, respectively.

## 2.4. Metal Doping

Previous studies have demonstrated the key role of electronegativity around the reaction domain in determining the dehydrogenation barriers of $MgH_2$.[21] Furthermore, over the course of almost a decade of experiments and calculations have demonstrated that the doping of transition metals can notably decrease the dehydrogenation temperature, hasten reaction kinetics, and greatly enhance the performance of hydrogen charging and discharging cycles.[18, 33-37] Herein, we further analyze the variation of dehydrogenation energy barriers when transition metal (TM) atoms are doped at the Mg site of $MgH_2$. Given the pronounced disparity in the -ICOHP values between Mg-H and TM-H bonds (*i.e.*, the bonding between TMs and H), coupled with the fact that the TM-H content is minimal in the experimental doped system and the primary reaction involves the rupture of Mg-H bonds, we are currently disregarding any dehydrogenation processes that do not involve Mg-H bonds. Ten 3*d* TM-doped $MgH_2$ models are constructed to analyze their electronegativity and neighborhood information, with the atomic H involved in the dehydrogenation process outlined in **Figure 5a**.

Since the analysis scope does not cover TM-H bonds, selecting an appropriate dehydrogenation path for doped structures is necessary. The initial and final structures of this path are shown in **Table S8** and **Figure S5**. Utilizing this dehydrogenation path, the overall electronegativity near the reaction area can be modified while disregarding the influence of the TM-H bonds on the analyzed structure. It is worth noting that the previously mentioned electronegativity applies not only to a single TM atom but also to single-atom catalysts that represent the overall environment of the graphene substrate (**Figure S6**), also known as the system electronegativity (**Eq. 10**):[21, 38]

$$X = (xX_C + yX_N - X_M) \times \frac{\theta_d}{n_d} \tag{10}$$

where $x$ and $y$ represent the number of carbon and nitrogen atoms of the catalytic active center, respectively. $X_C$, $X_N$, and $X_M$ represent the electronegativity of carbon, nitrogen, and the embedded single-atom metal, respectively. $\theta_D$ and $n_d$ represent the number of electrons present in the $d$ orbital and the maximum number of electrons that can exist, respectively. The primary influencing factor for the doped $MgH_2$ systems is the type of the metal atom, and the H vacancy information has already been corrected. Therefore, the electronegativity information for H will be disregarded in the subsequent model analyses.

From the analyzed structures, it is observed that the bond strength during the dehydrogenation process with TM dopants changes minimally. However, the size of the dehydrogenation barrier undergoes significant transformations (**Table S9**). On the one hand, this occurs because the reactive atomic H is situated far from the TM atom, resulting in a minimal change in the bonding strength. On the other hand, the atomic distance information and electronegativity exert dual influences, leading to considerable disparities in their respective dehydrogenation energy barriers. Remarkably, the results indicate that the dehydrogenation barrier of doped $3d$ TM atoms can fill in the voids of the calculated outcomes, thereby ensuring a seamless distribution of data points.

After several modifications and fitting procedures, the following electronegativity description information is designed, focusing on the region of the dehydrogenation reaction (wherein the value without TM doping is set to 1). The formula for the electronegativity neighborhood coefficient is given by **Eq. 11**:

$$\varphi_e = \frac{n_m \times X_{Mg}}{n_{TM} X_{TM} + n_{Mg} X_{Mg}} \tag{11}$$

where the electronegativity neighborhood coefficients are expressed as follows: $X_{Mg}$ and $X_{TM}$ represent the electronegativity of Mg and TM, respectively. $n_m$, $n_{TM}$, and $n_{Mg}$ represent the number of metal atoms, Mg, and TM atoms in the reaction center. It is noteworthy that the design philosophy of $\varphi_e$ is rooted in the impact of electronegativity on catalysis in heterojunction structures, making the descriptor applicable not only to doped systems but also to the systems with interface catalysis. In **Eq. 11**, five metal atoms adjacent to the surface of the dehydrogenation reaction are considered, and their positions are illustrated in **Figure 5a**. After modifying the description information of electronegativity, several reliable models are obtained with significant correlations (**Figure 5b-e** and **Table S10**). Specifically, the model utilizing the average value of all bond strengths and the two strongest bonds on either side of the reaction direction achieves high $R^2$ values (0.92 and 0.95, respectively).

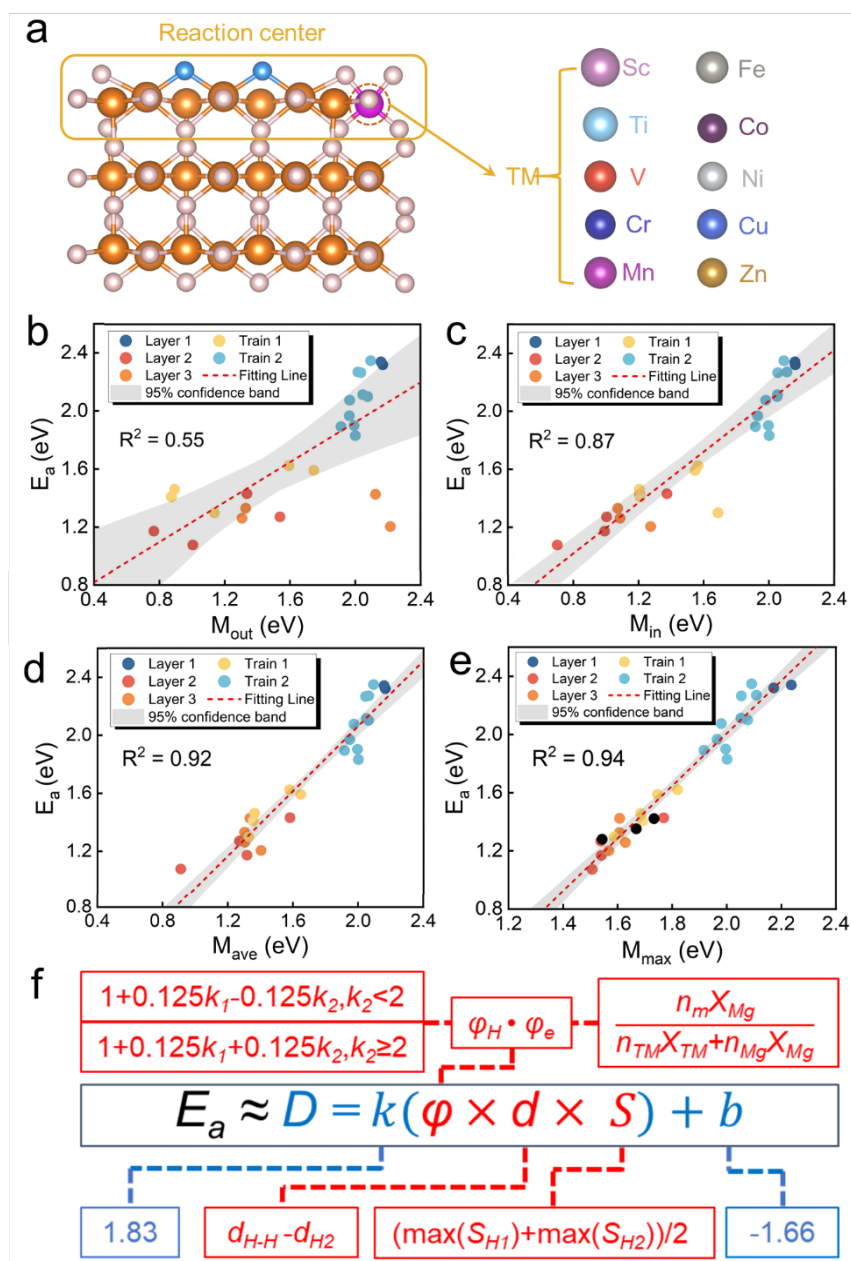

**Figure 5** Construction of electronic neighborhood coefficients for the $MgH_2$ dehydrogenation barrier descriptors. (a) Structural information of $MgH_2$ dehydrogenation to explore the effects of neighborhood electronegativity. White and orange spheres represent H and Mg, respectively. Blue spheres represent the participating atomic H in the reaction. (b-e) Model correlation analysis to modify electronegativity information: correlations between energy barriers and model values using (b) $S_{out}$, (c) $S_{in}$, (d) $S_{ave}$, and (e) $S_{max}$, respectively. (f) The derived model for describing dehydrogenation at $MgH_2$.

The above-constructed dehydrogenation barrier descriptors yield an optimal linear relationship with the dehydrogenation barrier over the $MgH_2$ surface. A descriptor that can more accurately depict the dehydrogenation barrier is formed following a simple linear refinement. The final composition of the

dehydrogenation model is shown in **Eqs. 12-15** and **Figure 5f**:

$$D_{max} = 1.83 M_{max} - 1.66 \tag{12}$$

$$M_{max} = \varphi_H \times \varphi_e \times d \times S_{max} \tag{13}$$

$$D_{max} = 1.83(1 + 0.125k_1 - 0.125k_2) \times \frac{n_m \times X_{Mg}}{n_{TM} X_{TM} + n_{Mg} X_{Mg}} \times (d_{H-H} - d_{H_2}) \times \frac{max(S_{H1\_out}, S_{H1\_in}) + max(S_{H2\_out}, S_{H2\_in})}{2} - 1.66, \; k_1 < 2 \tag{14}$$

$$D_{max} = 1.83(1 + 0.125k_1 + 0.125k_2) \times \frac{n_m \times X_{Mg}}{n_{TM} X_{TM} + n_{Mg} X_{Mg}} \times (d_{H-H} - d_{H_2}) \times \frac{max(S_{H1\_out}, S_{H1\_in}) + max(S_{H2\_out}, S_{H2\_in})}{2} - 1.66, \; k_1 \geq 2, \tag{15}$$

**2.4. Model Validation and Applications**

Afterward, three sets of out-of-sample dehydrogenation data are used to validate the model (**Table S11** and **Figure S7**). The initial structure and the desorbed atomic H are shown in **Figure 6a**. The energy barriers, bond strengths, H vacancies, and descriptor values for these three dehydrogenation pathways are provided in **Table S12**. The results after inserting the test points into the descriptor are shown in **Figure 6b**. Notably, **Tests 1-3** can be well-integrated into the descriptor, with the error margin as low as 0.04, 0.11, and 0.09 eV, respectively. Furthermore, the further inclusion of test data results in an improved correlation outcome compared to the previous results, achieving an $R^2$ of ~0.95. **Figure 6c** presents the accuracy of the descriptor by illustrating the prediction errors for all calculated data. As depicted, most of the prediction errors fall within 0.15 eV; the maximum error does not exceed 0.2 eV. These results demonstrate the strength and efficacy of the descriptor in capturing the complex dynamics of solid-state hydride dehydrogenation. The extended validation further supports its potential as a valuable tool for comprehending and anticipating dehydrogenation performance across various material systems.

The above results clearly show that the developed model not only closely matches the theoretical calculations but also aligns with the experimental conclusions from metal-doped $MgH_2$ materials. The energy barriers predicted by the descriptors show a high consistency with the onset dehydrogenation temperatures observed in some typical experiments reported to date (**Figure 6d** and **Table S13**).[39-44] Modifying the surface of $MgH_2$ can reduce the energy barrier for dehydrogenation, resulting in accelerated formation and growth of Mg nuclei on the surface for more efficient dehydrogenation and lower reaction temperature.[45, 46] Combining **Eqs.12-15**, **Figure 6d** demonstrates a clear negative correlation between the electronegativity of doped metals and dehydrogenation temperature. This finding is highly consistent with recent experimental studies.[47, 48] In addition, the metal doping ratio has a significant impact on the dehydrogenation temperature and the predicted energy barrier.

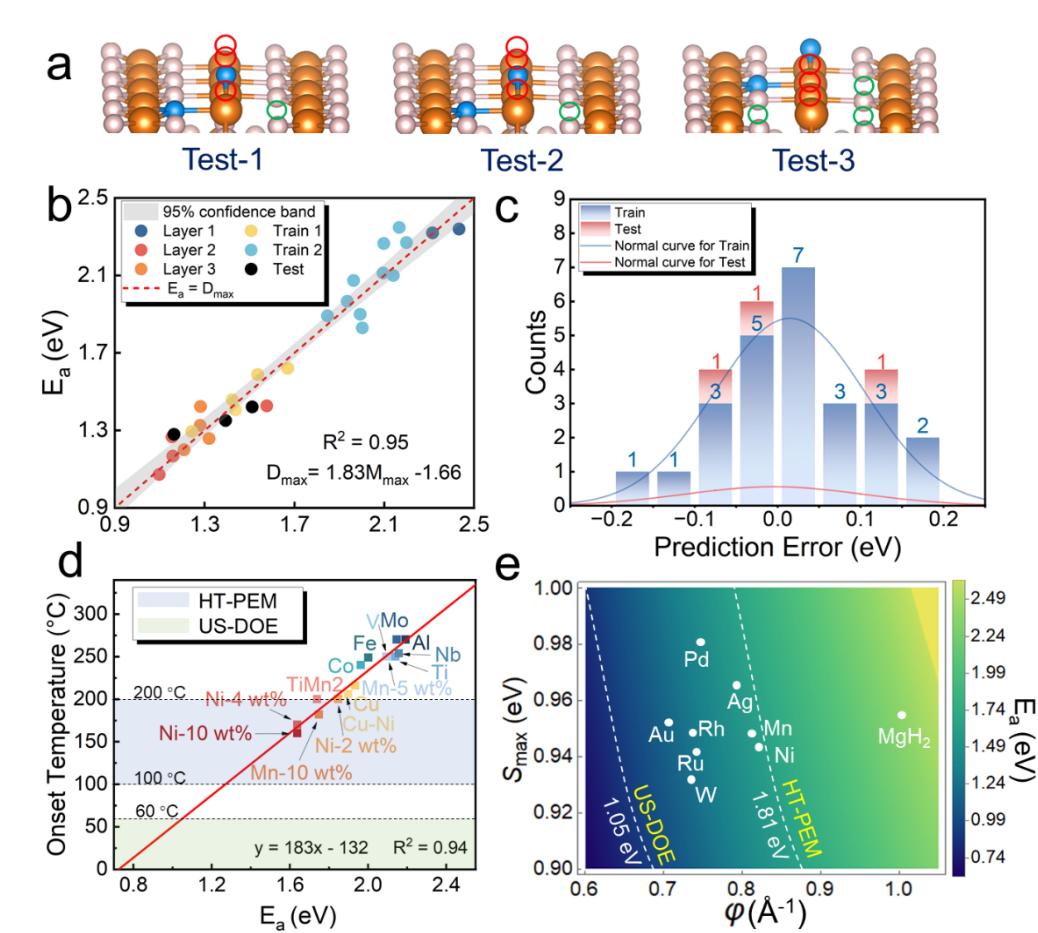

**Figure 6** Validating the accuracy of the derived model. (a) Structures of MgH$_2$ dehydrogenation for model validation. White and orange spheres represent H and Mg, respectively. Blue spheres and dashed circles represent the atomic H involved in the reaction and H vacancy, respectively. (b) Correlation between the values calculated from DFT+CI-NEB *versus* our model for 28 dehydrogenation barriers. (c) Statistical histograms of the prediction errors for out-of-sample data. The blue and yellow bars represent the number of errors present in the data used to train and test the model, respectively. (d) Correlation between onset dehydrogenation temperature and dehydrogenation barrier. (e) The derived contour plot predicts the dehydrogenation energy barrier of metal-doped MgH$_2$ as a function of $S_{max}$ and $\varphi$. The two white dashed lines represent the targets of the high-temperature proton exchange membrane (HT-PEM) and the US Department of Energy (US-DOE), respectively.

Currently, in the research of Mg-based hydrogen storage materials, only a few have been able to meet the on-board hydrogen storage temperature range (-40~60 °C) required by the US-DOE for light-duty fuel cell vehicles.[49] However, high-temperature proton exchange membrane (HT-PEM) fuel cells can operate between 100 to 200 °C.[50] According to our analysis, doping Mn and Ni may potentially meet the working temperature conditions, as their predicted energy barriers are lower than 1.81 eV. Since a small amount of TM-doping may lead to small changes in the surface atomic H distances (**Table S9**), the values of $\varphi$ and $S_{max}$ can be obtained through the relationship between experimental temperature and the descriptors, along with the model (**Figure 6e**). According to the electronegativity, it is predicted that various 4$d$ and 5$d$ metal dopants may lead to good dehydrogenation performance. However, from our theoretical predictions, we have not yet identified which element can achieve a dehydrogenation temperature below 60 °C through doping. Nevertheless, in experiments, a significant reduction in dehydrogenation temperature can be achieved through appropriate engineering technologies and the rational control of the dopant amount. Crucially, although the most promising solid-state hydrogen storage materials among doped metals have not yet fully met the objectives of the US-DOE, further application of this model to forecast performance enhancements through various forms of catalysis and nanotechnology will provide a swift approach to achieving the US-DOE's goals.

Furthermore, the model is primarily constructed from fundamental geometric distributions and elemental properties, ensuring its capability for formula expansion. By computing the kinetic energy barriers and bond states of various metal hydrides and recalibrating the coefficients and intercepts of our model, we can swiftly develop a descriptor model adept at forecasting the kinetic performance of metal hydrides across diverse structures, even including interstitial and complex hydrides. Subsequently, the model will swiftly predict and elucidate the influence of vacancies and doping on the hydrogen storage capacity of materials, thereby shortening the research cycle for understanding the mechanisms of hydrogen storage materials and speeding up the process of their modification and selection.

## 3. Conclusion

In summary, we have developed a data-driven model (**Figure 5f**) with the descriptors capable of accurately describing the dehydrogenation barriers of $MgH_2$. As shown in **Figure S8**, by progressively considering the Mg-H bond strength, H vacancy, atomic H distance, and metal doping, the accuracy of energy barrier predictions is enhanced. It was found that the average value of the strongest Mg-H bond strength of the -ICOHP of reactive atomic H provides the best description for the dehydrogenation barriers. This indicates that the dehydrogenation energy barrier is mainly determined by the region where atomic H has the strongest bonding with Mg.

Furthermore, the neighborhood coefficients of H vacancies were designed using the number of various types of H vacancies during the descriptor search. This not only improves the accuracy of the descriptor but also further reveals the mechanism by which the number of H vacancies on the two surfaces affects the energy barrier. Additionally, the descriptor range was expanded by doping TM atoms and proposing the electronegativity neighborhood coefficient. This coefficient shows a strong and positive correlation between the electronegative environment around the reaction and the dehydrogenation barrier.

Finally, three sets of out-of-sample data were compared with the predictions using the descriptors, with the errors of only 0.04, 0.11, and 0.09 eV, respectively. Moreover, it is noteworthy that the descriptor-based predictions of the constructed model not only closely match the outcomes of DFT, but also exhibit a striking resemblance to the temperature variations of $MgH_2$ dehydrogenation observed in the experiments involving metal doping. This leads us to conjecture that certain 4*d* and 5*d* transition metals, owing to their elevated electronegativity, may endeavor to further diminish the dehydrogenation temperature to align with the temperature standards set by the US-DOE. This model can accurately predict the dehydrogenation barriers of $MgH_2$ surface through the knowledge of Mg-H bond strengths. The effective descriptors, as the independent variables of the model, can expedite the analysis of $MgH_2$ dehydrogenation kinetics and reveal the primary influencing factors of surface dehydrogenation, which in turn can help design $MgH_2$-based materials with improved performance. Furthermore, the model's variables are not tied to Mg-based materials, enabling the formula to be adapted to various metal hydrides through rapid recalibration. We anticipate that this research will facilitate the discovery of novel metal hydride composite materials and solid-state hydrogen storage solutions.

**Code Availability**

For readers' convenience, the codes using the model to predict the dehydrogenation barriers are available *via*: https://github.com/Weijie-Yang/MgH2.

**Acknowledgments**

This work was funded by the Natural Science Foundation of Hebei Province of China（E2023502006）and Fundamental Research Fund for the Central Universities (2023MS125). H.L. acknowledges the Center for Computational Materials Science, Institute for Materials Research, Tohoku University for the use of MASAMUNE-IMR (No. 202312-SCKXX-0203) and the Institute for Solid State Physics (ISSP) at the University of Tokyo for the use of their supercomputers.

**Conflict of Interests**

There is no conflict of interest to declare.